\documentclass[prl,superscriptaddress,twocolumn]{revtex4-1}
\usepackage[colorlinks=true,urlcolor=blue]{hyperref}
\usepackage[dvips]{graphicx} 
\usepackage[english]{babel}
\usepackage{datetime}
\usepackage{verbatim}
\usepackage{amsmath}
\usepackage{siunitx} 
\usepackage{nccmath}

\usepackage{color}
\definecolor{red}{rgb}{0.75,0,0}
\definecolor{blue}{rgb}{0,0,0.75}
\definecolor{green}{rgb}{0,0.5,0}

\begin{document}

\widetext
 
\title{Geometric pinning and antimixing in scaffolded lipid vesicles}

\author{Melissa Rinaldin}
\affiliation{Instituut-Lorentz, Universiteit Leiden, P.O. Box 9506, 2300 RA Leiden, The Netherlands}
\affiliation{Huygens-Kamerlingh Onnes Lab, Universiteit Leiden, P. O. Box 9504, 2300 RA Leiden, The Netherlands}
\author{Piermarco Fonda}
\affiliation{Instituut-Lorentz, Universiteit Leiden, P.O. Box 9506, 2300 RA Leiden, The Netherlands}
\author{Luca Giomi}
\email{giomi@lorentz.leidenuniv.nl}
\affiliation{Instituut-Lorentz, Universiteit Leiden, P.O. Box 9506, 2300 RA Leiden, The Netherlands}
\author{Daniela J. Kraft}
\email{kraft@physics.leidenuniv.nl}
\affiliation{Huygens-Kamerlingh Onnes Lab, Universiteit Leiden, P. O. Box 9504, 2300 RA Leiden, The Netherlands}

\begin{abstract}


Unravelling the physical mechanisms behind the organisation of lipid domains is a central goal in cell biology and membrane biophysics. 
Previous studies on cells and model lipid bilayers featuring phase-separated domains found an intricate interplay between the membrane geometry and its chemical composition. However, the lack of a model system with simultaneous control over the membrane shape and conservation of its composition precluded a fundamental understanding of curvature-induced effects. 
Here, we present a new class of multicomponent vesicles supported by colloidal scaffolds of designed shape. We find that the domain composition adapts to the geometry, giving rise to a novel ``antimixed'' state. Theoretical modelling allowed us to link the pinning of domains by regions of high curvature to the material parameters of the membrane. 
Our results provide key insights into the phase separation of cellular membranes and on curved surfaces in general. 

\end{abstract}

\maketitle

Cells adapt the curvature of their membrane to create a plethora of functionalities from cargo-trafficking vesicles to cell division and motility. Dynamic remodelling of the membrane shape can be achieved by local changes in the lipid composition, the shape of the cytoskeleton, and the presence of curvature-generating proteins \cite{mcmahon2005membrane}. In turn, areas of high membrane curvature can recruit curvature-sensing proteins and lipids, which may further stabilise or increase local curvature differences \cite{peter2004bar}. 

Such a correlation between membrane geometry and chemical composition has also been extensively observed in model lipid membranes consisting of ternary mixtures of lipids and cholesterol. While being in a homogeneously mixed state at high temperatures, below a critical temperature, they separate into two different lipid phases. These phases, known as liquid-ordered (LO) and liquid-disordered (LD), differ in their lipid composition and possess different mesoscopic material properties such as layer thickness and resistance to bending. In cells, phase separation is ubiquitous and exploited to concentrate certain molecules \cite{Dolgin2018Nature}. Similar to cellular membranes \cite{rayermann2017hallmarks}, experiments on giant unilamellar vesicles (GUVs) \cite{baumgart2003imaging,baumgart2005membrane,hess2007shape,semrau2009membrane} and supported lipid bilayers (SLBs) \cite{Parthasarathy2006,subramaniam2010particle} reported a twofold correlation between these lipid domains and the membrane geometry: on the one hand, the local membrane curvature can favour the segregation of lipids and the nucleation and localization of domains. On the other hand, the presence of lipid domains can drive the formation of curved regions, such as buds, necks and protrusions. 

While these experiments have greatly contributed to our understanding of the physics and chemistry of lipid domains, they could not disentangle the correlation between membrane geometry and chemical composition. In GUVs the geometry cannot be constrained while the overall composition is conserved, whereas SLBs posses a fixed membrane geometry but also allow exchange of lipids with the environment. Furthermore, the change of the membrane curvature upon phase separation lead to repulsions between domains and non-equilibrium states. This prevented a fundamental understanding of curvature-related effects in phase-separating lipid membranes. Here we overcome these limitations by fabricating multicomponent scaffolded lipid vesicles (SLVs) which feature both prescribed shape and controllable composition. 

\begin{figure*}[t]
\includegraphics[width=1\linewidth]{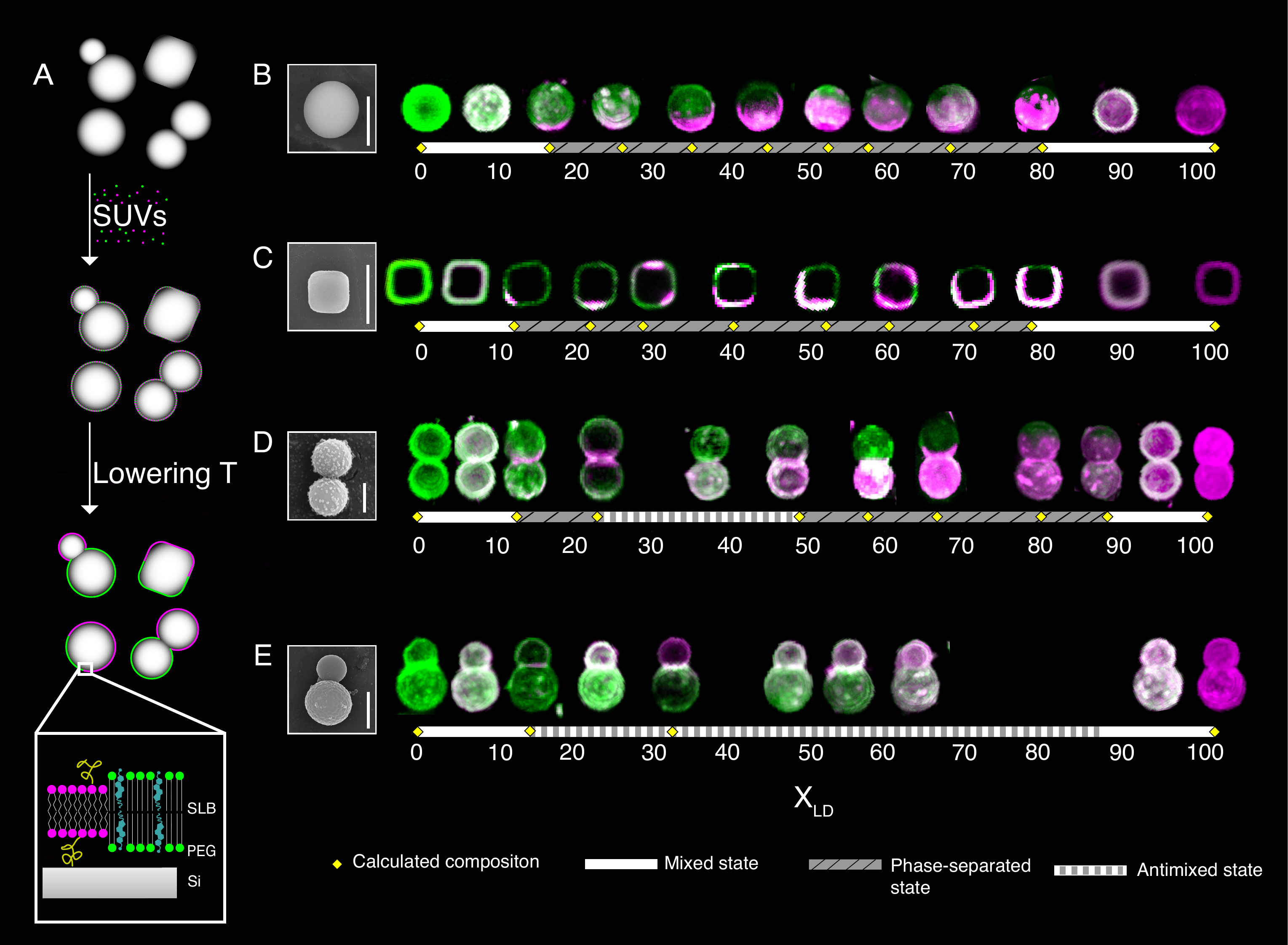}
\caption{\textbf{Experimental phase diagram of SLVs.} 
\textbf{A} Schematic representation of the experimental set-up: colloidal particles of spherical, cubic, dumbbell and snowman shape are coated with a lipid membrane at T=70 \si{\degreeCelsius}. After the temperature is lowered, the membrane undergoes phase separation. Liquid disordered (LD) and liquid ordered (LO) phases are represented in magenta and green, respectively. PEG molecules are drawn in yellow and cholesterol lipids in light blue. 
\textbf{B} Left: scanning electron microscopy (SEM) image of a spherical colloid. Right: top views of 3D reconstruction of spherical SLVs, ordered from left to right according to increasing percentage of LD area fraction. 
\textbf{C} Left: SEM image of a cubic colloid. Right: equatorial sections of SLVs on cubic scaffolds. 
\textbf{D} Left: SEM image of a dumbbell particle. Right: top views of 3D reconstructions of dumbbell-shaped SLVs. 
\textbf{E} Left: SEM image of a snowman colloid. Right: top views of 3D reconstructions of snowman-shaped SLVs. See Fig. S11 and Table 2 of the SI for respectively single channel images and area fraction values. Scale bars 2 \si{\micro\metre}.} \label{fig:one-set-up}
\end{figure*}

Our set-up, summarised in Fig. \ref{fig:one-set-up}A, consists of a ternary mixture of the lipids sphingomyelin (SM), 1-palmitoyl-2-oleoyl-\textit{sn}-glycero-3-phosphocholine (POPC) and cholesterol (chol) in a 2:1:1 mole ratio, deposited on micron-sized particles of four different shapes: spheres (Fig. \ref{fig:one-set-up}B), cubes (Fig. \ref{fig:one-set-up}C), symmetric dumbbells (Fig. \ref{fig:one-set-up}D) and asymmetric dumbbells (Fig. \ref{fig:one-set-up}E), also called snowman particles in the following. To preserve the fluidity of the lipid bilayer, we use colloidal particles with a silica surface and 5\% mole of 1,2-distearoyl-\textit{sn}-glycero-3-phosphoethanolamine-N-[methoxy(polyethylene glycol)-2000] (DOPE-PEG 2000) (see inset of Fig. \ref{fig:one-set-up}A). After lipid deposition, we lower the temperature to induce phase separation and image the SLVs using confocal microscopy. We identify the LO and LD phases through fluorescent labelling with N-[11-(dipyrrometheneboron difluoride)undecanoyl]-D-\textit{erythro}-sphingosylphosphorylcholine (C11 TopFluor SM), represented in green, and 1,2-dioleoyl-sn-glycero-3-phosphoethanolamine-N-lissamine rhodamine B sulfonyl 18:1 (Liss Rhod PE), represented in magenta \cite{almeida2003sphingomyelin, sezgin2012partitioning}. 

Remarkably, both the likelihood of phase separation and the spatial arrangement of the phase domains are deeply affected by the geometry of the substrate. By comparing samples of 200 SLVs of each type we found that only 22\% of the spheres showed phase separation. In contrast, for symmetric and asymmetric dumbbells, this fraction increased to 68\% and 91\%, respectively (SI: D.3). Since all experimental conditions were kept constant, we propose that the highly curved neck of the dumbbells promotes phase separation. This phenomenon is further enhanced in snowman particles, because of the different diameter, and hence curvature, of the two lobes. Although the chemical composition of SUVs is fixed prior to deposition, the actual SM:POPC:chol ratio on individual SLVs varies. This allows us to capture multiple points in the concentration phase space at once.

For successfully phase-separated SLVs, we observe a strong correlation between the curvature of the underlying scaffold and the area fractions, structure and location of the phase domains. We denote by $x_{\rm LO}$ and $x_{\rm LD}$ the relative area fractions occupied by the two liquid phases, so that $x_{\rm LO}+x_{\rm LD}=1$. We experimentally determined these area fractions by binarising the reconstructed profiles into the LO and LD phase (Methods: Imaging). Spherical scaffolds are uniformly curved, hence minimization of interface length 
between the LO and LD phases is the principal driving force of domain dynamics, which relaxes towards the systematic formation of two domains with relative area fractions set by the lipid composition (Fig. \ref{fig:one-set-up}B). In contrast to previous experiments \cite{Madwar2015a}, this observation implies that our system has reached equilibrium. 

Geometric effects start to become visible on cubic SLVs, where the curvature is predominantly localised at the corners and along the edges (see Fig. \ref{fig:one-set-up}C). At low $x_{\rm LD}$ (magenta), we predominantly observed the LD domains to be located at the edges, whereas at higher $x_{\rm LD}$ we could not identify a correlation with curvature. The regions of higher curvature localize the softer LD domains due to the lower cost of bending and we refer to this effect as \textit{geometric pinning}. 
This leads to the growth of multiple LD domains and hinders coalescence, although line tension still plays an important role in merging separate domains, similarly to the case of spherical SLVs. As a result, in a sample of 569 phase-separated cubes, $29\%$ showed three or more domains (see Fig. SI S13). 

In symmetric dumbbells and snowman particles geometric effects become more dramatic because of the presence of regions of negative (i.e. saddle-like) Gaussian curvature and high mean curvature. Fig. \ref{fig:one-set-up}D shows how the relative area fraction of the lipids determines the location of the interface in dumbbells. For low $x_{\rm LD}$ (below $\sim 23\%$), the softer disordered phase is pinned along the highly curved neck. For higher $x_{\rm LD}$ we observe only two domains: in the range $24-50\%$, the interface lies always along the neck, while for higher values it is positioned on one lobe. We find that $87\%$ of the 200 dumbbells exhibited an interface along the neck. This observation indicates that the energetic advantage resulting from positioning the interface along the neck is large enough to make any other configuration significantly unfavourable. Remarkably, the fact that the interface remains localised along the neck, for a whole range of concentrations, implies that the lipid composition of at least one of the two domains must vary. We name this unprecedented phenomenon \textit{antimixing}, to emphasise that the membrane composition is sharply inhomogeneous and yet features two coexisting {\em mixed} domains. 

The additional curvature asymmetry (radii ratio 0.57) in snowman particles increases this striking effect (Fig. \ref{fig:one-set-up}E). For $x_{\rm LD}$ equal to 13\%, three domains are present; above this value, we observed only two. In the latter case, the interface always lies along the neck and the compositions of each lobe vary continuously. Out of 200 snowman SLVs, $91\%$ showed antimixing. We conjecture that these unexpected results are a combination of three effects: the bending-rigidities-induced preference of one of the two phases for curved regions, the high stability of the interface provided by the neck region, and, crucially, the influence of curvature on the thermodynamics of phase separation. 

\begin{figure}
\begin{center}
\includegraphics[width=1\columnwidth]{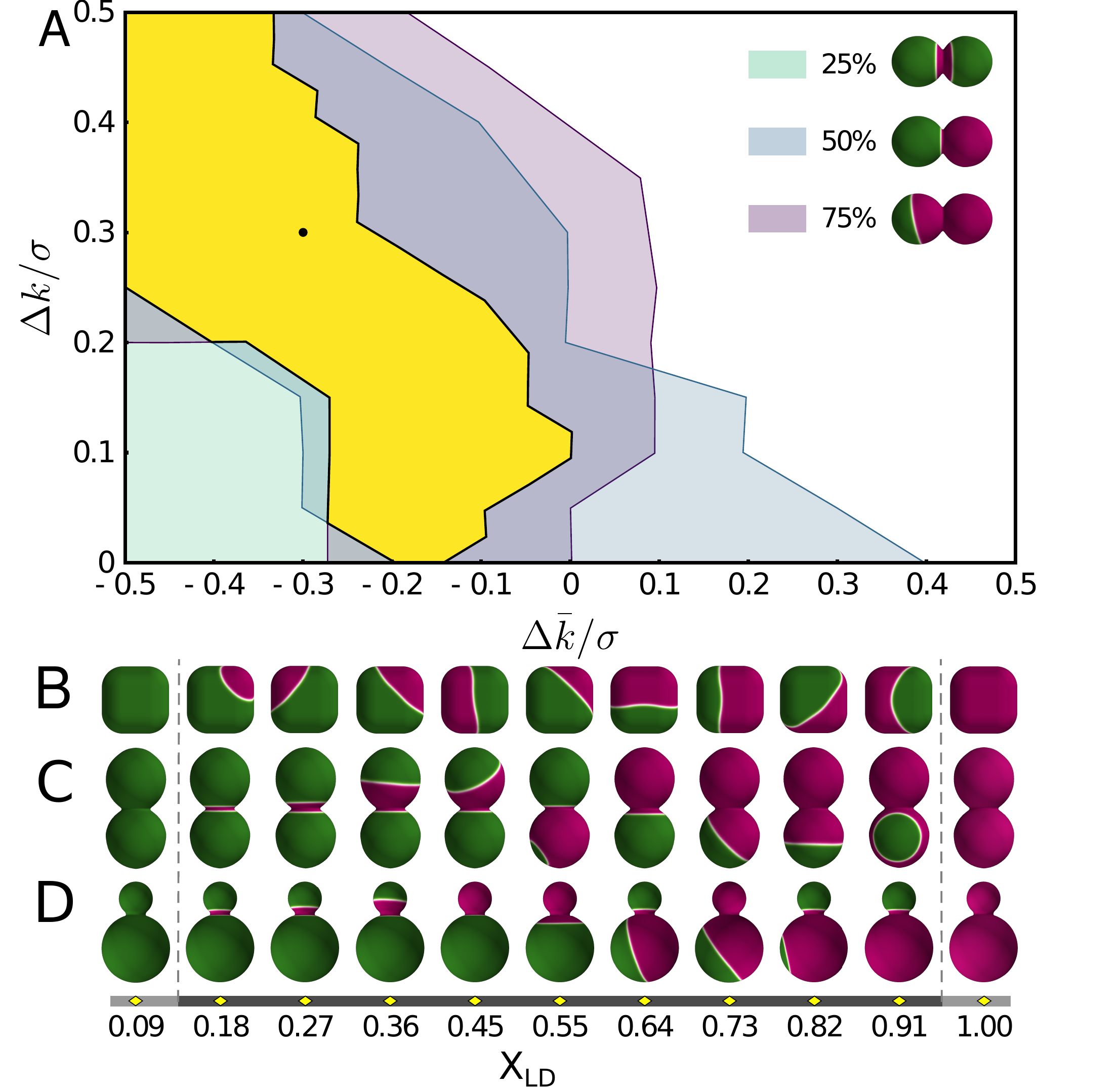}
\end{center}
\caption{
\textbf{Numerical simulation of phase domains.}
\textbf{A}
Allowed regions in the difference of bending rigidities parameter space where experimentally observed domain configurations can be reproduced through numerical simulations for a dumbbell-shaped membrane. The unit of length is such that the area of the dumbbell is equal to one. Shaded regions refer to the approximate portion of the phase space where observations can be reproduced, with each colour representing a different $x_{\rm LD}$ value, as indicated in the top-right inset. The yellow highlighted overlap region is the allowed parameter range for the bending rigidities, for fixed lateral line tension $\sigma$. The black dot in the highlighted regions corresponds to the point $(\Delta\bar{k},\Delta k)=(-0.3\sigma,0.3\sigma)$. \textbf{B-D} Equilibrium configurations for cubes, dumbbells and snowman particles for varying area fractions, using as coupling values the black dot in the top panel. We are unable to put upper bounds on $\Delta k$, nor to $\Delta \bar{k}$, and estimate that $\Delta k \simeq - \Delta \bar{k}$ in agreement with \cite{baumgart2005membrane}.
}
\label{fig:figure2}
\end{figure}

To shed light on the experimental results, we have modelled the membrane as a closed surface $\Sigma$ endowed with a continuous order parameter $\phi$ representing the local concentration of unsaturated lipids, so that $\int_{\rm \Sigma} {\rm d}A\,\phi$ corresponds to the area occupied by the LD phase. The equilibrium configurations of $\phi$ are determined by minima of the following free-energy inspired by J\"ulicher-Lipowsky theory of multicomponent vesicles \cite{Julicher1993,elliott2010surface}:
\begin{equation}
F
= 
\int_\Sigma {\rm d}A\,
\left[
\frac{\epsilon}{2} | \nabla \phi |^2 + \frac{1}{\epsilon} V(\phi) + k(\phi) H^2 + \bar{k}(\phi) K
\right]\;,
\label{eq:energy}
\end{equation}
where $H$ and $K$ are the mean and Gaussian curvatures of $\Sigma$. The gradient term in the energy favours gently-varying field configurations, while the second term, $V(\phi)= V_{0}\phi^2 (\phi-1)^2$, is a double-well potential that favours phase-separated configurations, where $\phi$ is uniform over portions of $\Sigma$ which we identify as the LO and LD phases. The parameter $\epsilon$ controls the thickness of the one-dimensional interfaces separating the two phases. In the limit $\epsilon \to 0$ the first two terms of Eq. \eqref{eq:energy} converge to the interfacial energy of Ref. \cite{Julicher1993} with line tension $\sigma=\frac{2}{3}\sqrt{2V_{0}}$ (see SI \S C.2). The functions $k(\phi)$ and $\bar{k}(\phi)$ represent the elastic moduli of the LO and LD phases and can be parametrised as $k(\phi)=\Delta k\,f(\phi)$ and $\bar{k}(\phi)=\Delta\bar{k}\,f(\phi)$, with $\Delta k$ and $\Delta \bar{k}$ the difference in bending rigidity and saddle-splay modulus between the LO and LD phase and $f(\phi)$ a dimensionless function interpolating between 0 and 1. Here we choose $f(\phi)=\phi^{2}(3-2\phi)$, but the precise form of this function is uninfluential. Finally, mass conservation is enforced via a Lagrange multiplier $\lambda$, so that the total energy is given by $F+\lambda \int_\Sigma {\rm d}A\,\phi$.

Free-energy minimisation in the sharp interface limit (i.e. $\epsilon\rightarrow 0$) yields:
\begin{equation}
\sigma \kappa_g = \Delta k \, H^2 + \Delta \bar{k} \, K + \lambda \,,
\label{eq:interface}
\end{equation}
where $\kappa_g$ is the geodesic curvature of the interface (see SI \S C.1). Eq. \eqref{eq:interface} is the two-dimensional analogue of the Young-Laplace equation for liquid-liquid interfaces: on a flat substrate, where both $H$ and $K$ vanish, solutions describe a circular droplet of radius $\sigma/\lambda$, with the Lagrange multiplier effectively working as a Laplace pressure across the interface. For spherical particles of radius $R$, for which $H^{2}=K=1/R^2$, the equilibrium interface lies along constant geodesic curvature lines, i.e non-maximal circles. The solid angle underlying such circles is either fixed by mass conservation ($ \lambda \neq 0$) or by the elastic moduli differences of the two phases. On more general curved substrates, the effective Laplace pressure is affected by the geometric properties of the environment, modulated by the difference in rigidity between the two lipid phases. 

Although intuitive, Eq. \eqref{eq:interface} is extremely difficult to solve for more complex shapes, such as those considered in our experiments. However, one can construct approximated solutions of Eq. \eqref{eq:interface} by numerical minimisation of the free energy Eq. \eqref{eq:energy} for finite $\epsilon$ (see SI \S C.3). Fig. \ref{fig:figure2}A (inset) shows three examples for symmetric dumbbells having $x_{\rm LD}$ equal to $25\%$, $50\%$ and $75\%$ corresponding to the 4th, 6th and 9th configuration of Fig. \ref{fig:one-set-up}D. By comparison with the experimental results we identified the region of appropriate values for the material parameters $\Delta k/\sigma$ and $\Delta \bar{k}/\sigma$, indicated by the yellow hue in Fig. \ref{fig:figure2}. We found good agreement for $\Delta k >0$, $\Delta \bar{k}<0$ and $|\Delta k| \simeq |\Delta \bar{k}| \simeq 10^{-1} \sigma$ (for units of length normalised with the area of the particle), consistent with previous estimates for free-standing bilayers \cite{baumgart2005membrane}. Choosing the values $\Delta k = 0.3 \sigma$ and $\Delta \bar{k}=-0.3 \sigma$ from the region of appropriate parameters identified for dumbbells, we then calculated the equilibrium configurations for cubic and snowman shaped particles (Fig. \ref{fig:figure2}B-D). By comparing Fig. \ref{fig:one-set-up}C and Fig. \ref{fig:figure2}B, we see that the model qualitatively reproduces the observed phase-separation patterns on cubic scaffolds. Similarly, Fig. \ref{fig:one-set-up}D and Fig. \ref{fig:figure2}C show good agreement for most area fractions on dumbbell particles. On the other hand, comparing Figs. \ref{fig:one-set-up}E and \ref{fig:figure2}D, we find that antimixing cannot be captured by the sharp interface limit of Eq. \eqref{eq:energy}, as expected.

\begin{figure} 
\begin{center}
\includegraphics[width=1\columnwidth]{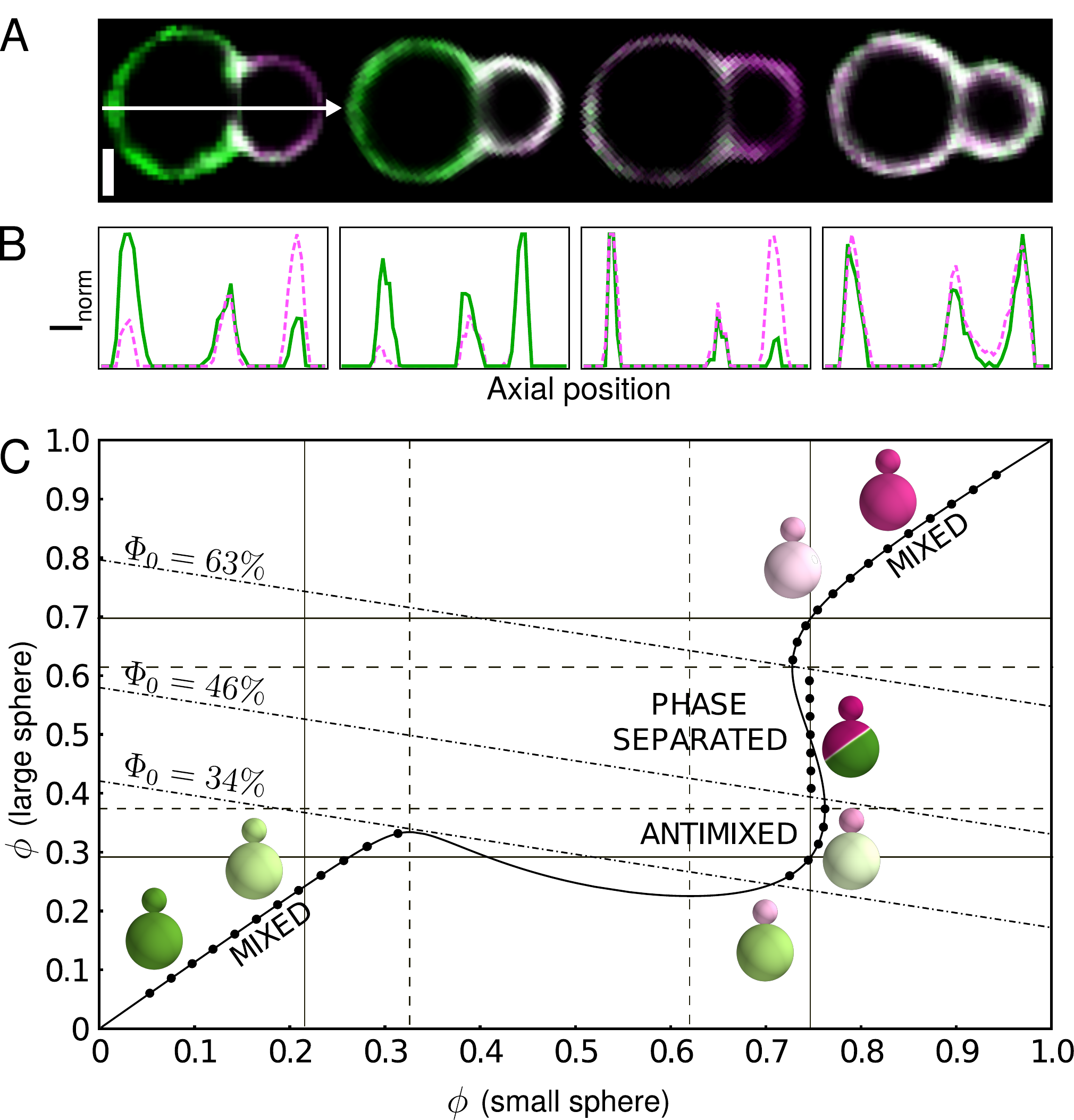} 
\end{center}
\caption{
\textbf{The antimixed state. A} From left to right: snowman shaped SLVs in phase-separated state, antimixed states and mixed state. Scale bar 1 \si{\micro\metre}. \textbf{B} Normalised intensity profiles along the axial direction of the TOP Fluor SM and DOPE Rhodamine in green and magenta respectively. Phase-separated SLVs show different intensity peaks in both lobes, while mixed configurations exhibit overlap of the whole intensity profile. Antimixed SLVs display overlap of peaks in one lobe and different peaks in the other. \textbf{C} Minimal model of antimixing on snowman particles. The particles are approximated by two disjoint spheres allowed to exchange lipids. The diagram shows the local concentration $\phi$ of unsaturated lipids on each sphere, with the diagonal dot-dashed lines indicating the configurations of equal average concentration $\Phi_{0}=\frac{1}{A}\int {\rm d}A\,\phi$. The vertical and horizontal dashed (solid) lines mark the spinodal (binodal) point on each sphere. The solid curve crossing the diagram diagonally represents the line of equilibria obtained by analytically minimising the free energy. When such a line is in between the spinodal lines, the lipids are phase-separated. The black dots are computed from numerical simulations and match the analytical values outside the spinodal region. (Inset) 3D reconstruction of representative configurations. The colouring of mixed and antimixed configurations is pure green for $\phi=0$ and pure magenta for $\phi=1$. Lipids are antimixed for $34\% \leq \Phi_0 \leq 46\%$ and phase-separated for $46\% \leq \Phi_0 \leq 63\%$.
\label{fig:figure3}
}
\end{figure}

To shed light on the origin of antimixing in snowman particles (Fig. \ref{fig:figure3}A,B), we have developed a simplified variant of the model, where the effect of curvature on the thermodynamics of phase separation is explicitly taken into account. Snowman particles are approximated by two disjoint spheres which are allowed to exchange lipids. The neck region acts as a barrier for lipid domains to migrate from one lobe to the other, but does not stop single lipids to diffuse freely. Then, in Eq. \eqref{eq:energy} we take $V(\phi)= J \phi (1-\phi) + \phi \log \phi + (1-\phi) \log (1-\phi)$, with $J$ a constant. Such a thermodynamic potential, corresponding to the mean-field free-energy for a lattice-gas system, can account for mixed and phase-separated configurations alike and reduces to the usual double-well potential in the presence of phase-separation. Finally, we set $k(\phi)+\bar{k}(\phi)=J/\epsilon\,f(\phi)$, so that the curvatures can affect directly the thermodynamic potential arbitrarily close to the thin interface limit (i.e. $\epsilon\rightarrow 0$, SI \S C.4). Minimising the free energy for fixed average concentration, $\Phi_{0}=\frac{1}{A}\int {\rm d}A\,\phi$, yields the phase diagram of Fig. \ref{fig:figure3}C, inclusive of the mixed, antimixed and phase-separated state. We stress here that $\Phi_{0}$ equates the area fraction $x_{\rm LD}$ only in the case of phase-separated configurations, whereas for mixed and antimixed configurations it corresponds to the relative amount of unsaturated lipids. The mixed and antimixed configurations form a {\em line of equilibria} (solid line) for which the concentration $\phi$ is uniform across each lobe, but differs between the two lobes in case of antimixed configurations. The latter scenario does not occur if lipid mixing is unaffected by the curvature of the substrate (in which case $\phi=\Phi_{0}$ and the line coincides with the bisectrix of the diagram). We therefore conclude that the observed antimixed state of Fig. \ref{fig:one-set-up}E and Fig. \ref{fig:figure3}A,B originates mainly from the fact that the geometry of the substrate affects not only the lateral organisation of lipid domains but also the free-energy landscape of the order parameter. 
Though this phenomenon shares some similarity with the curvature-driven lipid sorting observed by Sorre {\em et al.} \cite{sorre2009curvature}, it does not require our system to be on the verge of demixing and is a system-wide effect. 

In this article we have introduced a new experimental model system to decipher the complex physics of phase separation on curved surfaces and cells in particular: scaffolded lipid vesicles (SLVs). Like the cytoskeleton and proteins shape the cellular membrane, SLVs impose their geometry onto the supported lipid membrane allowing an independent study of geometry-induced effects. With this unique setup, we discovered a novel state of lipid membranes, the antimixed state, in which the composition of the lipid domains adapts to the geometry of the  scaffold. The antimixed state occurs when the underlying geometry is highly inhomogeneous, and leads to coexisting mixed phases, whose concentrations of saturated and unsaturated lipids adapts to the overall composition.

While the interplay between geometry and localization of the domains had been previously observed, setting the scaffold geometry allowed us to fully understand the delicate balance between the curvature of the surface, the material parameters and the composition of the membrane that leads to geometric pinning. Surprisingly, there is an upper and a lower bound for the area fraction of the softer phase for geometric pinning to occur, whose precise values are determined by the scaffold shape. Our results on geometric pinning and antimixing could be extended to other physical systems where phase separation occurs \cite{franzmann2018phase}. They provide key insights into how cells might regulate local concentrations of lipids and proteins by adapting the membrane shape. 
The here developed membrane-coated colloids can furthermore switch their surface properties and, upon further chemical functionalization, their interactions \cite{chakraborty2017colloidal}, from uniform to site-specific, paving the way towards smart materials and biomedical applications such as sensing, imaging and drug delivery. 

\paragraph*{Acknowledgements} 
This work was supported by the Netherlands Organisation for Scientific Research (NWO/OCW), as part of the Frontiers of Nanoscience program (MR), VENI grant 680-47-431 (DJK) and the VIDI scheme (LG,PF). We thank Vera Meester for help with particle synthesis and electron microscopy imaging. 

\paragraph*{Author contributions} 
MR and DJK designed the experiment. MR performed the experiments and the data analysis. PF and LG developed the theoretical model. PF performed the analytical and numerical calculations. LG and DJK conceived the study and supervised the research. All authors wrote the article.

\bibliography{references}{}

\clearpage 
\newpage

\section*{Materials and Methods} 

\subsection*{Reagents}
1-palmitoyl-2-oleoyl-sn-glycero-3-phosphocholine (POPC), porcine brain sphingomyelin (Brain SM), ovine wool cholesterol, 1,2-dioleoyl-sn-glycero-3-phosphoethanolamine-N-lissamine rhodamine B sulfonyl 18:1 (Liss Rhod PE), N-[11-(dipyrrometheneboron difluoride)undecanoyl]-D-erythro-sphingosylphosphorylcholine (C11 TopFluor SM), 1,2-dioleoyl-sn-glycero-3-phosphoethanolamine-N-[methoxy(polyethylene glycol)-2000] (DOPE PEG 2000) were purchased from Avanti Polar Lipids. 

Silica spheres were purchased from Microparticles GmbH (2.06 $\pm$ 0.05 \si{\micro\metre} and $7 \pm 0.29$ \si{\micro\metre}). 
Polystyrene-3-(Trimethoxysilyl)propyl methacrylate (PS-TPM) dumbbell and snowman particles were synthesised by making a protrusion from swollen polystyrene particles \cite{kim2006synthesis}. 
Hematite cubic particles were made following the method of \cite{sugimoto1993formation} and coated with silica \cite{Rossi2011}.
Details about particle syntheses in SI \S A.1.

Hepes buffer was made with 115 mM NaCl, 1.2 mM $\mathrm{CaCl}_2$, 1.2 mM $\mathrm{MgCl}_2$, 2.4 mM $\mathrm{K_2HPO_4}$ and 20 mM Hepes. 

\subsection*{Lipid bilayer coating}
A mixture of 500 \si{\micro\gram} of SM, POPC and cholesterol in 2:1:1 mole ratio was prepared in chloroform. 0.2\% mole fraction of Liss Rhod PE and C11 TopFluor SM were added to fluorescently label the liquid-disordered and liquid-ordered phases respectively. 5\% mole fraction of DOPE PEG 2000 was added to improve the mobility of the bilayer. The lipids were dried in two hours by vacuum desiccation for two hours and then re-suspended to a 2 mg/mL solution with Hepes buffer. The solution was vortexed for 15 minutes and heated to 70 \si{\degreeCelsius}. The lipid solution was extruded 21 times with a mini extruder (Avanti Polar Lipids) equipped with two 250 \si{\micro\gram} gas-tight syringes (Hamilton), four drain discs and nucleopore track-etch membranes (Whatman) and placed on a heating plate set at 70\si{\degreeCelsius}. Then, in a rotavapor (Buchi) 50 \si{\micro\liter} of SUVs were added to 1 mL of 0.05 \%w/v of particles dispersed in Hepes. Particles were left undergoing gentle rotation at 70\si{\degreeCelsius} for one hour. The temperature and the time were chosen to keep the SUVs in the mixed state during the spreading and to give them enough time to spread on the surface of the particles. The solution was then centrifuged at 3000 rpm for 10 minutes and the supernatant replaced with Hepes to remove any SUVs in excess.
\\

\subsection*{Imaging and analysis}
The SLVs were imaged at room temperature with an inverted confocal microscope (Nikon Eclipse Ti-E) equipped with a Nikon A1R confocal scanhead with galvano and resonant scanning mirrors. A 100x oil immersion objective (NA=1.4) was used. 488 and 561 nm lasers were used to excite excited respectively Top Fluor and Lissamine Rhodamine dyes. Lasers passed through a quarter wave plate to avoid the polarisation of the dyes and the emitted light was separated into two channels using a dichroic mirror equipped with 500-550 nm and 565-625 nm filters.
Mobility of the bilayer was check by fluorescence recovery after photobleaching (FRAP) experiments in which the recovery of the signal was fitted with an exponential curve (SI \S A.2). Three-dimensional image stacks were acquired by scanning the sample in the z direction with a MCL Nano-drive stage and reconstructed with Nikon AR software. \\
The composition of the SLVs was analysed by using Python scripts. For spherical, dumbbell and snowman particles every slice of the z-stack was fitted with a sphere, a nephroid and a two circles. For cubic particles only the equatorial plane was fitted with a squircle. The intensity along the fitted shapes was normalized by the maximum intensity and binarised, to obtain the ratio between the liquid ordered and disordered phases. For z-stacks this ratio was then averaged along the z-direction to obtain the total ratio of area fraction between the two phases, that is reported in Figure \ref{fig:one-set-up}. The scripts can be found on \href{https://github.com/RinaldinMelissa/SLVsAreaFraction}{Github} (https://github.com/RinaldinMelissa/SLVsAreaFraction) and a more detailed analysis description of the analysis method is reported in SI \S A.4.

\end{document}